\documentclass[10pt]{article}
\usepackage{amsmath,amssymb}
\usepackage{graphicx}
\usepackage{cite}

\numberwithin{equation}{section}

\begin{document}

\title{Charged perfect fluids in the presence of a cosmological constant}

\author{
C. G. B\"ohmer\footnote{c.boehmer@ucl.ac.uk}~\mbox{} and
A. Mussa\footnote{atifahm@math.ucl.ac.uk}\\[1ex]
Department of Mathematics and Institute of Origins\\ 
University College London, Gower Street \\
London, WC1E 6BT, United Kingdom
}

\maketitle

\begin{abstract}
We consider the static and spherically symmetric field equations of general
relativity for charged perfect fluid spheres in the presence of a cosmological
constant. Following work by Florides (1983) we find new exact solutions of the
field equations, and discuss their mass radius ratios. These solutions, for
instance, require the charged Nariai metric to be the vacuum part of the
spacetime. We also find charged generalizations of the Einstein static universe
and speculate that the smallness problem of the cosmological constant might
become less problematic if charge is taken into account.
\end{abstract}

\section{Introduction}

Ever since Schwarzschild constructed the first exact interior solution of the
Einstein field equations, static and spherically symmetric matter solutions have
been of great interest. Additional degrees of freedom of the matter were
included at later stages, like the cosmological constant, anisotropic pressure
and charge. An entire new class of interior Schwarzschild like solutions was
obtained by Florides~\cite{Florides:1974}, these particular solutions were
distinct since they featured a vanishing radial stress $T^r_r$. This
Schwarzschild like solution was later generalized to include a non-zero charge
distribution, see~\cite{Florides:1977,Mehra:1982,Gron:1986}. In each case the
new interior solution was matched to an exterior Reissner-Nordstr\"om solution,
or the exterior Schwarzschild solution in the absence of charge. 

Static charged fluid spheres had been studied in detail by Efinger (1965), for
instance, who was interested in understanding charged particles in general
relativity, see
also~\cite{Bonnor:1960,Efinger:1965,Wilson:1967,Kramer:1971,Bekenstein:1971,
Bailyn:1972,Gron:1986}. Note that the spacetimes obtained so far were not
singularity free. Then Mehra (1979) managed to construct an interior solution
and avoid a singularity. Florides (1983) also considered charged perfect fluid
solutions, these solutions described the complete field, interior and exterior.
The neutral Florides interior solution was then extended further
by~\cite{Xu:1986ka} to include the effects of a non zero cosmological constant
$\Lambda$. This solution was matched to the exterior Schwarzschild-de Sitter
(Kottler) solution. Since then not much attention has been devoted to analyze
charged perfect solutions in the presence of a cosmological constant.

In this paper we follow~\cite{Florides:1974} and impose the restriction that the
fluid component of the matter is isotropic. In this way we obtain new classes of
interior charged solutions in the presence of a cosmological constant. A
particular class of these solutions, namely the charged analogue of the interior
Nariai solution is discussed in greater detail. We show how to match this
solution to the exterior charged Nariai solution. Note that the exterior is not
asymptotically flat and thus has a somewhat special place in the family of
Reissner-Nordstr\"om de Sitter solutions. We are also able to construct charged
generalizations of the Einstein static universe where the pressure is
monotonically decreasing or increasing between the two regular centers. 

\section{Field equations}

We consider the system of Einstein-Maxwell equations with cosmological term
\begin{align*}
  &G_{ij} + \Lambda g_{ij} = 8\pi T_{ij}\,, \qquad 
  T_{ij} = M_{ij} + E_{ij}\,,\\
  &\partial_j(\sqrt{-g}F^{ij}) = \sqrt{-g} J^{i}\,, \qquad
  \partial_{[i}F_{jk]}=0\,,
\end{align*}
where we set $G=c=1$. The matter energy-momentum tensor is denoted by $M_{ij}$,
and the electromagnetic energy-momentum tensor $E_{ij}$ is given by
\begin{align*}
  E_{ij} = \frac{1}{4\pi} \left(
  F_{i}{}^{k}F_{jk}-\frac{1}{4}g_{ij}F^{mn}F_{mn}\right)\,.
\end{align*}
The 4-current density vector $J^i$ is defined by
\begin{align*}
  J^i = 4\pi \sigma u^i\,.
\end{align*}
The proper charge density is denoted by $\sigma$ and $u^i$ is the 4-velocity
satisfying $u^i u_i = -1$.

Following the notation of~\cite{Florides:1983}, we consider the static and
spherically metric in the form
\begin{align}
  ds^2 = -e^{a(r)} dt^2 + e^{b(r)} dr^2 + r^2 d\Omega^2\,,
  \label{eqn:e}
\end{align}
where $d\Omega^2 = d\theta^2 + \sin^2\negmedspace\theta\, d\phi^2$ is the line
element of the unit sphere.

Using this metric ansatz and the above defined 4-current density vector, the
only non-vanishing component of the electromagnetic tensor $F_{ij}$ becomes
\begin{align}
  F_{rt} = \frac{e^{(a+b)/2}}{r^2} q\,, \qquad
  q = 4\pi \int_0^r \sigma e^{b/2} r'^2 dr' \,.
\label{eqn:sth}
\end{align}
Therefore, the resulting components of $E_i^j$ are
\begin{align}
  E_t^t = E^r_r = -E^\theta_\theta = -E^\phi_\phi =
  -\frac{q^2}{8\pi r^4}\,.
  \label{eqn:h}
\end{align}
As for the matter we choose an anisotropic perfect fluid
\begin{align}
  M_i^j = \mbox{diag}(-\rho,p,p_{\perp},p_{\perp})\,.
  \label{eqn:i}
\end{align}

The resulting $(tt)$ and $(rr)$ components of the Einstein-Maxwell field
equations and the energy-momentum conservation equation take the following form
\begin{align}
  \frac{1}{r^2} \frac{d}{dr}\left(r-re^{-b} \right) - \Lambda &=
  8\pi \rho + \frac{q^2}{r^4}\,,
  \label{eqn:field1}\\
  \frac{e^{-b}}{r^2} + \frac{a' e^{-b}}{r} - \frac{1}{r^2}
  + \Lambda &= 8\pi p - \frac{q^2}{r^4}\,,
  \label{eqn:field2}\\
  p' + \frac{a'}{2}(\rho+p)+\frac{2}{r}(p-p_{\perp})
  - \frac{qq'}{4\pi r^4} &= 0\,.
  \label{eqn:field3}
\end{align}
Therefore, this system consists of three independent equations for six unknown
functions $\{a,b,\rho,p,p_{\perp},q\}$. Therefore the system is under-determined
and additional physical or mathematical assumptions need to be taken into
account in order to close the set of equations.

For example, in the neutral case $q \equiv 0$, with an isotropic perfect fluid
$p=p_{\perp}$ there are three equations and four unknowns. In that case the most
physical approach is to consider a barotropic equation of state for the matter,
$\rho=\rho(p)$ which closes the set of equations. Another popular approach
motivated by astrophysical observations is the prescription of a density profile
$\rho(r)$. However, this usually leads to a singular center and is therefore
problematic from a rigorous point of view.

In the charged case, one can impose various physical conditions to decrease the
number of unknown functions. One can for instance assume that the total energy
momentum tensor is isotropic
\begin{align}
  T^r_r = T^\theta_\theta = T_\phi^\phi\,,
  \label{eqn:j}
\end{align}
which yields the condition
\begin{align}
  p - p_{\perp} = \frac{1}{4\pi}\frac{q^2}{r^4}\,.
  \label{eqn:k}
\end{align}
Therefore, this assumption leads to a relation between the anisotropy
(difference between radial and tangential pressure) and the electromagnetic
field. Note that in the absence of charges this immediately leads to an
isotropic perfect fluid.

The second natural assumption is to consider an isotropic perfect fluid
\begin{align}
  M^r_r = M^\theta_\theta = M_\phi^\phi\,,
  \label{eqn:l}
\end{align}
which simply gives
\begin{align}
  p = p_{\perp}\,.
  \label{eqn:m}
\end{align}
Note that more conditions need to be imposed to close the system of equations.

A useful linear combination is the sum of equations~(\ref{eqn:field1}) and~(\ref{eqn:field2})
which yields
\begin{align}
  a' + b' = 8\pi re^b (\rho + p)\,.
  \label{eqn:help}
\end{align}

The first field equation~(\ref{eqn:field1}) can be formally integrated and we obtain
\begin{align}
  e^{-b} &=
  1 - \frac{1}{r}\int_0^r\left(8\pi\rho +
  \frac{q^2}{r'^4}\right)r'^2 dr' -\frac{\Lambda}{3}r^2
  \nonumber \\ &=
  1 - \frac{2m_g}{r} + \frac{q^2}{r^2} - \frac{\Lambda}{3}r^2\,,
  \label{eqn:o}
\end{align}
where we used~(\ref{eqn:sth}) and where
\begin{align}
  m_g = m_i + m_q &= \int_0^r 4\pi \rho r'^2 dr'
  + \int_0^r 4\pi r' \sigma q e^{b/2} dr'\,.
  \label{eqn:p}
\end{align}

The field equations can be combined into the useful Tolman-Oppenheimer-Volkoff
equation, which is a differential equation for the pressure expressed in terms
of the other matter quantities. Eliminating $a'$ from~(\ref{eqn:field3})
using~(\ref{eqn:field2}) and~(\ref{eqn:o}), we find
\begin{align}
  p' = -r \frac{(4\pi p + \frac{m_g}{r^3} -
\frac{q^2}{r^4}-\frac{\Lambda}{3})(\rho+p)}{1-\frac{2m_g}{r}+\frac{q^2}{r^2}
-\frac{\Lambda}{3}r^2} -
\frac{2}{r}(p-p_\perp) - \frac{qq'}{4\pi r^4}\,,
\end{align}
which reduces to the well-known Tolman-Oppenheimer-Volkoff equation when $q \equiv 0$,
$\Lambda = 0$ and $p \equiv p_\perp$.

If the interior solution is of finite extent, we denote total quantities with capital letters, this means $M,Q$ denote the total mass and total charge of the solution, respectively. We use $r_b$ to denote the boundary of the object, defined as the vanishing pressure surface. Therefore, if the exterior spacetime is the Reissner-Nordstr\"om de Sitter spacetime, its metric is given by
\begin{multline}
  ds^2 = -\left(1 - \frac{2M}{r} + \frac{Q^2}{r^2} - \frac{\Lambda}{3}r^2\right)
dt^2 \\+ \left(1 -
\frac{2M}{r} + \frac{Q^2}{r^2} - \frac{\Lambda}{3}r^2\right)^{-1} dr^2 + r^2
d\Omega^2\,,
\end{multline}
where $r \geq r_b$.

The main aim of this work is to find new exact solutions of the field equations.
Therefore, in view of Eq.~(\ref{eqn:o}) one particular choice would be to
consider the case when
\begin{align}
  8\pi\rho + \frac{q^2}{r^4} = 8\pi \mu \equiv \mbox{constant}\,.
  \label{eqn:q}
\end{align}
This condition has been first considered in~\cite{Coop:1978} where a charged
dust was considered. It has further been exploited by~\cite{Florides:1983}. In
this case the interior metric takes the simpler form 
\begin{align}
  e^{-b} &= 1 - \frac{r^2}{R^2}\,,\label{eqn:aaa}\\
  \frac{1}{R^2} &= \frac{8\pi\mu}{3} + \frac{\Lambda}{3}\,.
  \label{eqn:s}
\end{align}
Combining equations~(\ref{eqn:s}) and~(\ref{eqn:aaa}) with (\ref{eqn:q}) yields the metric function $e^{-b}$ in the form
\begin{align}
  e^{-b} = 1 - \frac{8\pi\rho}{3} r^2 - \frac{q^2}{3 r^2} - 
  \frac{\Lambda}{3} r^2\,.
  \label{eqn:s1}
\end{align}

\section{Solutions of the field equations}

\subsection{Florides' solution}

Let us consider an isotropic total energy-momentum tensor (Condition 1) which
gives~(\ref{eqn:k}). Inserting this into the conservation
equation~(\ref{eqn:field3}) gives
\begin{align}
  p' + \frac{1}{2}a'(\rho+p) + \frac{1}{2\pi}\frac{q^2}{r^5} -
\frac{1}{4\pi}\frac{qq'}{r^4} = 0 \,.
\end{align}
Now we substitute $a'$ using~(\ref{eqn:help}) and use the fact that
\begin{align*}
  -\frac{d}{dr}\left(\frac{q^2}{8\pi r^4}\right) = \frac{1}{2\pi}\frac{q^2}{r^5}
  -\frac{1}{4\pi}\frac{qq'}{r^4}\,,
\end{align*}
we arrive at the equation
\begin{align*}
  p' - \frac{1}{2}b'(\rho+p) + 4\pi r e^b (\rho+p)^2 
  -\frac{d}{dr}\left(\frac{q^2}{8\pi r^4}\right) = 0 \,.
\end{align*}
Finally, we introduce the quantity $z= \rho + p$ which results in 
\begin{align}
  z' - \frac{1}{2}b'z + 4\pi r e^b z^2  -\frac{d}{dr}\left(\rho +
  \frac{q^2}{8\pi r^4}\right) = 0 \,.
\end{align}

If we assume the term in the bracket to be a constant (Condition 2), see the
condition~(\ref{eqn:q}), then this equation simplifies considerably which we
will assume henceforth, which yields
\begin{align}
  z' - \frac{1}{2} b' z + 4\pi r e^b z^2 = 0\,.
  \label{eqn:t}
\end{align}
Next, making use of~(\ref{eqn:aaa}) we have
\begin{align}
  b' = \frac{2r/R^2}{1-r^2/R^2} \,,
  \label{eqn:u}
\end{align}
and therefore
\begin{align}
  z' - \frac{r/R^2}{1-r^2/R^2} z + \frac{4\pi r}{1 - r^2/R^2} z^2 = 0\,,
  \label{diff}
\end{align}
which can be integrated using separation of variables
\begin{align}
  z = \frac{1}{4\pi R^2}\biggl[1 + C\sqrt{1-\frac{r^2}{R^2}}\biggr]^{-1}\,,
  \label{eqn:w}
\end{align}
where $C$ is a constant of integration. 

The other metric function $a$ follows from~(\ref{eqn:help})
\begin{align}
  e^{a/2} = \frac{1+C\sqrt{1-\frac{r^2}{R^2}}}{1+C}\,,
  \label{eqn:x}
\end{align}
where we fixed the other constant of integration by requiring $e^a(r=0)=1$.
Therefore we found the complete metric. Note that in a static spacetime we can
always rescale the time coordinate arbitrarily.

\subsection{Towards new solutions}

In~\cite{Florides:1983}, the constant of integration $C$ was determined by the
condition that the pressure vanishes at some radius $p(r_b)=0$. Since this
condition requires the existence of a vanishing pressure hypersurface, we follow
a different approach and fix the constant of integration using the central
values of pressure and energy density.

Let us define the central pressure and energy density to be $p_c=p(r=0)$ and
$\rho_c=\rho (r=0)$, respectively. Then
\begin{align}
  \rho_c + p_c = z_c(r=0) = \frac{1}{4\pi R^2(1+C)}\,,\\
  C = -1 + \frac{1}{4\pi R^2(\rho_c + p_c)}\,.
  \label{eqn:y}
\end{align}

In the chosen coordinate system there is a coordinate singularity when $r
\rightarrow R$. The spatial part of the metric is a 3-sphere, hence it is
natural to introduce the third Euler angle of the sphere by
\begin{align}
  r = R \sin\chi\,, \qquad 1 - \frac{r^2}{R^2} = \cos^2 \chi\,.
  \label{eqn:z}
\end{align}
In these new coordinates the interior metric takes the form
\begin{align}
  ds^2 = - \Bigl(\frac{1+C\cos\chi}{1+C}\Bigr)^2 dt^2 + R^2 d\chi^2 +
  R^2 \sin^2\negmedspace \chi d\Omega^2,
  \label{eqn:aa}
\end{align}
which covers the entire 3-sphere. Note that the matter part of the solution does
not necessarily have to `fill' this three sphere. In fact, solutions analogous
to the Schwarzschild interior solution have their vanishing pressure surface
located before the equator of this three sphere and thus this new coordinate is
not important in understanding them. However, in the presence of a cosmological
constant one can construct solutions where the matter part occupies more than
half the three sphere. Understanding these solutions requires a proper
coordinate system covering the entire space.

Let us now insert the constant of integration expressed as central values into
the interior metric
\begin{multline}
  ds^2 = - \Bigl(4\pi R^2 (\rho_c + p_c)(1-\cos\chi) + \cos\chi \Bigr)^2 dt^2
\\+ R^2 d\chi^2 + R^2
\sin^2\negmedspace \chi d\Omega^2\,.
  \label{eqn:bb}
\end{multline}
While we have been able to find the metric, in order to find all unknown functions, we have to impose one additional condition to close the system. As this final condition (Condition 3) we assume $q = e r^2 = e R^2 \sin^2\negmedspace \chi$, where $e$ has dimension $\rm{charge/length^2}$. This condition immediately implies that the effective energy density is constant, by virtue of~(\ref{eqn:q}). Note that the resulting energy momentum tensor is regular everywhere, however, the proper charge density $\sigma$ blows up like $1/r$ near the center by virtue of~(\ref{eqn:sth}).

This allows us to find the radial  pressure of the matter configuration as a
function of the radius or the third Euler angle. For instance, using $p = z -
\rho$ we obtain
\begin{align}
  p=\frac{
(\rho_c+p_c)(4\pi\rho_c-\Lambda-e^2)-\rho_c(4\pi(\rho_c+3p_c)-\Lambda-e^2)
\cos\chi}{(4\pi(\rho_c+3p_c)-\Lambda-e^2)\cos\chi-12\pi(\rho_c+p_c)}\,.
  \label{pressure}
\end{align}
This pressure function vanishes at $\chi=\chi_b$ where
\begin{align}
  \cos\chi_b =
  \frac{(\rho_c+p_c)(4\pi\rho_c-\Lambda-e^2)}{
  \rho_c(4\pi(\rho_c+3p_c)-\Lambda-e^2)}\,.
\end{align}
Note that $\chi_b$ does not have to exist. In fact, we will construct solutions
where the pressure never vanishes.

Since we have found an expression of the form $\chi_b = \chi_b(\rho_c,p_c,\Lambda,e^2)$, we can also express the central pressure in terms of the other quantities. For this we find
\begin{align}
  p_c =
\frac{\rho_c(1-\cos\chi_b)(4\pi\rho_c-\Lambda-e^2)}{
4\pi\rho_c(3\cos\chi_b-1)+\Lambda+e^2}\,.
  \label{cpressure}
\end{align}
Since we are only interested in matter distributions with regular center, we assume the central pressure to be finite which is equivalent to the positivity of the denominator of~(\ref{cpressure}). Therefore, we find
\begin{align}
  \cos\chi_b > \frac{1}{3} - \frac{\Lambda}{12\pi\rho_c} -
  \frac{e^2}{12\pi\rho_c}\,,
  \label{chi}
\end{align}
which is a generalization of the well-known Buchdahl inequality. To see this, we go back to the radial coordinate $r$ and use that $\cos(\arcsin x) = \sqrt{1-x^2}$. This gives
\begin{align}
  \cos\left(\arcsin\frac{r_b}{R}\right) = \sqrt{1-\frac{r_b^2}{R^2}} = e^{-b(r_b)/2}\,,
  \label{eqn:n1}
\end{align}
by virtue of~(\ref{eqn:aaa}). Our Condition 3 states that $q/r^2 = e$, inserting this into~(\ref{eqn:s1}) leads to
\begin{align}
  e^{-b(r_b)} = 1 - \frac{8\pi\rho}{3} r_b^2 - \frac{e^2}{3} r_b^2 - 
  \frac{\Lambda}{3} r_b^2\,.
  \label{eqn:n2}
\end{align}
Let us finally denote $M=4\pi/3\, \rho r_b^3$, then equation~(\ref{chi}) becomes
\begin{align}
  \sqrt{1-\frac{2M}{r_b}-\frac{e^2}{3}r_b^2-\frac{\Lambda}{3}r_b^2} > \frac{1}{3} -
\frac{\Lambda}{12\pi\rho_c} - \frac{e^2}{12\pi\rho_c}\,,
\end{align}
which is the familiar form of the Buchdahl inequality, see also~\cite{Boehmer:2007gq,Andreasson:2008xw}. Letting $\Lambda,e^2 \rightarrow 0$ gives the textbook result
\begin{align}
  \sqrt{1-\frac{2M}{r_b}} > \frac{1}{3}\,, \qquad \frac{2M}{r_b} < \frac{8}{9}\,.
\end{align}

\subsection{Classes of new solutions}

Based on the pressure function~(\ref{pressure}) we can straightforwardly
identify new classes of solutions. Firstly, let us assume that the pressure
vanishes at the equator of the three sphere, this means when $\chi_b=\pi/2$. This
type of solution occurs when the relation $\Lambda + e^2 = 4\pi\rho_c$ holds.
When $\Lambda + e^2 > 4\pi\rho_c$, the pressure vanishes after the equator
of the three sphere. If we furthermore want the pressure to vanish before the
second center of the three sphere, we must impose $4\pi\rho_c < \Lambda + e^2 <
8\pi\rho_c (\rho_c+2p_c)/(2\rho_c+p_c)$.

Next, we can encounter situation where the pressure does not vanish anywhere and
the solution has a second regular center, this means we assume the pressure at
the second center to be finite. In that case $8\pi\rho_c
(\rho_c+2p_c)/(2\rho_c+p_c)<\Lambda + e^2<8\pi(2\rho_c+3p_c)$. If $\Lambda$
exceeds this upper limit the pressure diverges. In this class of solutions,
there is one special solution where the pressure is constant. This happens when
$\Lambda+e^2 = 4\pi(\rho_c+3p_c)$ which is a charged generalization of the
Einstein static universe. In the following Section, we will discuss these new
types of solutions in detail.

\section{New exact solutions}

\subsection{Solutions with exterior Nariai metric}

Let us assume $\Lambda + e^2 = 4\pi\rho_c$. In this case the pressure
function~(\ref{pressure}) vanishes when $\chi_b=\pi/2$, the equator of the three
sphere. Now we need to identify the correct vacuum solution matched at the
vanishing pressure surface. It is clear that the vacuum solution in this case
cannot be part of the Reissner-Nordstr\"om de Sitter spacetime. The only other
static and spherically symmetric spacetime with charge and cosmological constant
is the charged Nariai metric, we collect some basic facts of this solution in the
Appendix,~\ref{appsec}.

This metric is given by
\begin{align}
  ds^2 = \frac{1}{A}\Bigl(-(\alpha \sin\psi + \beta \cos \psi)^2 dt^2 +
d\psi^2\Bigr) + \frac{1}{B}
d\Omega^2\,,
  \label{eqn:na}
\end{align}
where the cosmological constant and the total charge are related to $A$ and $B$
by
\begin{alignat}{2}
  \Lambda&=\frac{1}{2}(A+B)\,, &\qquad Q^2&=\frac{1}{2}(B-A)\,,\\
  A&=\Lambda-Q^2\,, &\qquad B&=\Lambda+Q^2\,.
\end{alignat}
This is the most general form of the charged Nariai solution, the constants
$\alpha$ and $\beta$ can be chosen arbitrarily and will be fixed by our matching
conditions. In the following we will show that metric~(\ref{eqn:na}) is the
correct exterior for~(\ref{eqn:aa}) or~(\ref{eqn:bb}) when the matching is
performed at the $\chi=\pi/2$ hypersurface.

There are two equivalent approaches to matching two metrics at a given
hypersurface. One can either match the first and second fundamental forms or one
can introduce Gauss coordinates relative to the hypersurface and then show that
metric is continuous and differentiable at the matching surface. We will follow
the latter approach since both metrics are almost in the correct form. 

Let us introduce a new coordinate $\zeta = R \chi$ for~(\ref{eqn:aa}), which we
will call the interior metric
\begin{align}
  ds^2_{\rm int} = -\Bigl(\frac{1+C \cos(\zeta/R)}{1+C}\Bigr)^2 dt^2 + 
  d\zeta^2 + R^2 \sin^2(\zeta/R) d\Omega^2\,.
\end{align}
Let us also introduce a new coordinate $\zeta = \psi/\sqrt{A}$ for
metric~(\ref{eqn:na}), the exterior metric
\begin{align}
  ds^2_{\rm ext} = -\frac{1}{A}(\alpha \sin(\sqrt{A}\,\zeta) + \beta
\cos(\sqrt{A}\,\zeta))^2 dt^2
+ d\zeta^2 + \frac{1}{B} d\Omega^2\,.
\end{align}
The matching surface (vanishing pressure surface) is located at $\zeta = \pi
R/2$.

Continuity of the metric yields
\begin{align}
  R^2 &= \frac{1}{B} \\
  \frac{1}{1+C} &= \frac{1}{\sqrt{A}}(\alpha \sin(\sqrt{A}\,\pi R/2) + \beta
  \cos(\sqrt{A}\,\pi R/2))
\end{align}
while continuity of the first derivative implies
\begin{multline}
  -\frac{2}{1+C}\frac{C}{R} = \frac{2}{A}(\alpha \sin(\sqrt{A}\,\pi R/2) + \beta
\cos(\sqrt{A}\,\pi
R/2))\\ \times (\alpha\sqrt{A} \cos(\sqrt{A}\,\pi R/2) - \beta\sqrt{A}
\sin(\sqrt{A}\,\pi R/2))\,.
\end{multline}
It is straightforward to solve these equations for $\alpha$ and $\beta$. The
most elegant approach seems to be to firstly use that $R=1/\sqrt{B}$, and
secondly use the second equation for the first factor of the right-hand side of
the third. This gives
\begin{align}
  \sin\Bigl(\frac{\sqrt{A}}{\sqrt{B}}\frac{\pi}{2}\Bigr) \alpha +
\cos\Bigl(\frac{\sqrt{A}}{\sqrt{B}}\frac{\pi}{2}\Bigr) \beta &=
\frac{\sqrt{A}}{1+C}\,,\\
  \cos\Bigl(\frac{\sqrt{A}}{\sqrt{B}}\frac{\pi}{2}\Bigr) \alpha -
\sin\Bigl(\frac{\sqrt{A}}{\sqrt{B}}\frac{\pi}{2}\Bigr) \beta &= -\frac{2C}{R}\,.
\end{align}
In the absence of charge, $A=B=\Lambda$ and these equations simplify
considerably since the trigonometric functions become either $1$ or $0$. Let us
denote $\xi = \sqrt{A/B} \pi/2$, then one can write the latter equations as the
following simple linear system of equations
\begin{align}
  \begin{pmatrix} \sin\xi & \cos\xi \\ \cos\xi & -\sin\xi \end{pmatrix}
  \begin{pmatrix} \alpha \\ \beta \end{pmatrix} = 
  \begin{pmatrix} \sqrt{A}/(1+C) \\ -2C/R \end{pmatrix}\,.
\end{align}
The matrix on the left is its own inverse and thus $\alpha$ and $\beta$ are
given by
\begin{align}
  \begin{pmatrix} \alpha \\ \beta \end{pmatrix} = 
  \begin{pmatrix} \sin\xi & \cos\xi \\ \cos\xi & -\sin\xi \end{pmatrix}
  \begin{pmatrix} \sqrt{A}/(1+C) \\ -2C/R \end{pmatrix}\,.
\end{align}

Therefore, we have shown that the interior and the exterior metric can be
matched, the metric being $C^1$ at the surface. Without further assumptions,
this cannot be improved, as can be seen by the following argument. The energy
density inside the charged star is constant, while it is zero in the vacuum
region. Therefore, the energy-momentum tensor has a jump. By virtue of the
Einstein-Maxwell field equations, the Einstein tensor must have a jump too.
Since it contains the second derivatives of the metric, the metric is at most
$C^1$.

We would also like to note that these interior solutions can be interpreted as matter solutions which require the Bertotti-Robinson spacetime~\cite{Bertotti:1959,Robinson:1959} to be the electro-vacuum part of the manifold if we assume a vanishing cosmological constant. 

\subsection{Solutions with black hole event horizons}

If $4\pi\rho_c < \Lambda + e^2 < 8\pi\rho_c (\rho_c+2p_c)/(2\rho_c+p_c)$, then
the pressure vanishes after the equator of the three sphere but before its
second center. In this case the vanishing pressure surface is in a region where
the area group orbits are decreasing. The vacuum part of this spacetime is part
of the Reissner-Nordstr\"om de Sitter solution. However, this region contains
the black hole event horizon. This unusual class of solutions has been studied
in the past in the absence of charge, see~\cite{Boehmer:2002gg,Boehmer:2003uz}.
Since these solutions appear to have little physical relevance, we will not
discuss them further.

\subsection{Generalized Einstein static universes}

We now assume that $\Lambda + e^2$ is in the range
$8\pi\rho_c(\rho_c+2p_c)/(2\rho_c+p_c)<\Lambda + e^2<8\pi(2\rho_c+3p_c)$. In
this case the pressure is strictly positive. The spacetime has two regular
centers and it is natural to refer to such spacetimes as generalizations of the
Einstein static universes. 

Within these solutions, there is the charged generalization of the original
Einstein static universe. If we choose $\Lambda+e^2 = 4\pi(\rho_c+3p_c)$, the
prefactors of the trigonometric functions in the pressure~(\ref{pressure})
vanish and the pressure is constant. The radius of this charged Einstein static
universe is $1/R^2 = 4\pi (\rho_c + p_c)$. 

It should be noted that in all of the above solutions, the cosmological constant
and the charge parameter appear in the combination $\Lambda + e^2$ and one can
regard this as an effective cosmological constant, note that $e^2 \geq 0$. For
the charged Einstein static universe in particular this means that we can
construct such a matter configuration without cosmological constant. In this
case the charge acts as an external force that pulls the matter apart and
balances the gravitational attraction. 

Solutions with two centers are symmetric with respect to $\chi = \pi/2$. To see
this, let us denote the pressures at the two centers by $p_1$ and $p_2$,
respectively, then
\begin{align}
  p_1 &:= p(\chi = 0) = p_c \,,\\
  p_2 &:= p(\chi = \pi) = \frac{2\rho_c(4\pi\rho_c - \Lambda -
e^2)+p_1(16\pi\rho_c-\Lambda
-e^2)}{\Lambda+e^2 - 8\pi(2\rho_c +3p_1)}\,.
\end{align}
The latter equation can be solved for $\Lambda + e^2$, which then becomes a
function of $\rho_c,p_1,p_2$
\begin{align}
  \Lambda + e^2 = \frac{8\pi\rho_c(\rho_c+2p_1+2p_2)+3p_1p_2}{2\rho_c+p_1+p_2}\,.
\end{align}
This allows us to eliminate $\Lambda + e^2$ from the pressure
function~(\ref{pressure}) and we arrive at
\begin{align}
  p(\chi)[\rho_c,p_1,p_2] = \frac{\rho_c(p_1+p_2) + 2p_1 p_2 +
\rho_c(p_1-p_2)\cos\chi}{2\rho_c +
p_1  + p_2 - (p_1-p_2)\cos\chi}\,.
  \label{press2}
\end{align}
The pressure vanishes at $\chi_b$ where
\begin{align}
  \cos\chi_b =
\frac{(\rho_0-p_1)(p_2(2p_1-1)-p_1)}{p_1(\rho_0+p_1)+p_2(2\rho_0p_1-\rho_0+p_1)}
\,.
\end{align}

The aforementioned symmetry property of the pressure function can now easily be
expressed mathematically, namely we have
\begin{align}
  p(\pi/2-\gamma)[\rho_c,p_1,p_2] = p(\pi/2+\gamma)[\rho_c,p_2,p_1]\,.
\end{align}

\section{Conclusions}

The aim of this paper was to construct new classes of exact solutions of
Einstein's field equations in a static and spherically symmetric setting. We
considered a charged perfect fluid to be the matter source. One class of new
solutions is an interior solution which requires the charged Nariai metric to
describe the vacuum part of this spacetime. The other class generalizes the
Einstein static universe. These solutions are characterized by two regular
centers and a non-uniform pressure function~(\ref{press2}). They possess an
additional symmetry property, namely, the pressure is point-symmetric with
respect to the middle between the two centers.

In our particular setting the cosmological constant $\Lambda$ and the charge $e$
always appear in the combination $\Lambda + e^2$ which we can either view as an
effective charge or an effective cosmological constant. It is the latter point
of view which might be of interest in the context of cosmology. If these results
are not an artefact of our setting, then a small overall charge density
throughout the universe might be viewed as a natural explanation of a small
cosmological constant.

\section*{Acknowledgements}

AM would like to thank Filipe Mena for useful discussions, and the organisers of
GR19 where parts of this work were presented.

The authors also wish to express their gratitude to the referees of this paper for their constructive reports.

\appendix

\section{The charged Nariai solution}
\label{appsec}

The charged Nariai solution forms a two dimensional subspace of the
Reissner-Nordstr\"om de Sitter class of solutions. The inner, outer and
cosmological horizons are all equivalent, sometimes called degenerate. Note that
this implies
\begin{align}
  1-\frac{2M}{r} + \frac{Q^2}{r^2} -\frac{\Lambda r^2}{3}=0\,,
\end{align}
the breakdown of the the $(r,t)$ coordinate system which describes only part of
the complete spacetime. We will follow the approach in~\cite{Ortaggio:2002}
where it is noted that the Nariai spacetime can be viewed as a 4 dimensional
submanifold of a flat 6 dimensional Lorentzian manifold with product structure.
Let us consider
\begin{align}
  ds^2 = -dx_0^2 + dx_1^2 +dx_2^2 + dx_3^2 + dx_4^2 + dx_5^2\,,
  \label{app}
\end{align}
such that
\begin{align}
  -x_0^2 + x_1^2 + x_2^2 = \frac{1}{A}\,,\qquad x_3^2 + x_4^2 +x_5^2 =
\frac{1}{B}\,,
\end{align}
with $A \neq B = {\rm const.}$ and $A+B = 2 \Lambda$. Thus this spacetime is a
direct product of two 3 dimensional manifolds, one being a 3 sphere and the
other one being a hyperbolic 3 space. To make this relation explicit, let us
consider the following parametrization 
\begin{align}
  x_0 &=
  \frac{1}{\sqrt{A}}\frac{1}{\sqrt{\alpha^2 + \beta^2}}(\alpha \sin\psi + \beta
\cos\chi) 
  \sinh(\sqrt{\alpha^2 + \beta^2} t)\,, \nonumber\\
  x_1 &=
  \frac{1}{\sqrt{A}}\frac{1}{\sqrt{\alpha^2 + \beta^2}}(\alpha \sin\psi + \beta
\cos\chi) 
  \cosh(\sqrt{\alpha^2 + \beta^2} t)\,, \nonumber\\
  x_2 &=
  \frac{1}{\sqrt{A}}\frac{1}{\sqrt{\alpha^2 + \beta^2}}(\beta \sin\psi - \alpha
\cos \chi)\,,
  \nonumber\\
  x_3 &=
  \frac{1}{\sqrt{B}}\sin\theta \cos\phi\,,
  \nonumber\\
  x_4 &=
  \frac{1}{\sqrt{B}}\sin\theta \sin\phi\,,
  \nonumber\\
  x_5 &=
  \frac{1}{\sqrt{B}}\cos\theta\,.
\end{align}
Metric~(\ref{app}) then gives metric~(\ref{eqn:na}).

\begin{thebibliography}{99}

\bibitem{Andreasson:2008xw}
H.~Andreasson.
\newblock {Sharp bounds on the critical stability radius for relativistic
  charged spheres}.
\newblock {\em Commun. Math. Phys.}, 288:715--730, 2008.

\bibitem{Bailyn:1972}
M.~{Bailyn} and D.~{Eimerl}.
\newblock {General-Relativistic Interior Metric for a Stable Static Charged
  Matter Fluid with Large e/m}.
\newblock {\em Phys. Rev.}, D5:1897--1907, 1972.

\bibitem{Bekenstein:1971}
J.~D.~{Bekenstein}.
\newblock {Hydrostatic Equilibrium and Gravitational Collapse of Relativistic
  Charged Fluid Balls}.
\newblock {\em Phys. Rev.}, D4:2185--2190, 1971.

\bibitem{Bertotti:1959}
B.~{Bertotti}.
\newblock {Uniform Electromagnetic Field in the Theory of General Relativity}.
\newblock {\em Phys. Rev.}, 116:1331--1333, 1959.

\bibitem{Boehmer:2002gg}
C.~G.~B{\"o}hmer.
\newblock General relativistic static fluid solutions with cosmological
  constant.
\newblock 2002.
\newblock {unpublished Diploma thesis}.

\bibitem{Boehmer:2003uz}
C.~G.~B{\"o}hmer.
\newblock Eleven spherically symmetric constant density solutions with
  cosmological constant.
\newblock {\em Gen. Rel. Grav.}, 36:1039--1054, 2004.

\bibitem{Boehmer:2007gq}
C.~G.~B{\"o}hmer and T.~Harko.
\newblock {Minimum mass-radius ratio for charged gravitational objects}.
\newblock {\em Gen. Rel. Grav.}, 39:757--775, 2007.

\bibitem{Bonnor:1960}
W.~B.~{Bonnor}.
\newblock The mass of a static charged sphere.
\newblock {\em {Zeitschrift f\"ur Physik}}, 160:59--65, 1960.

\bibitem{Coop:1978}
F.~I.~{Cooperstock} and V.~{de La Cruz}.
\newblock {Sources for the Reissner-Nordstrom metric}.
\newblock {\em Gen. Rel. Grav.}, 9:835--843, 1978.

\bibitem{Efinger:1965}
H.~J.~{Efinger}.
\newblock {{\"U}ber die Selbstenergie und Ladung eines durch
  Gravitationswirkungen stabilisierten Teilchens in einem gekr{\"u}mmten Raum}.
\newblock {\em {Zeitschrift f\"ur Physik}}, 188:31--37, 1965.

\bibitem{Florides:1974}
P.~S.~{Florides}.
\newblock {A New Interior Schwarzschild Solution}.
\newblock {\em {Proceedings of the Royal Society A}}, 337:529--535, 1974.

\bibitem{Florides:1977}
P.~S.~{Florides}.
\newblock {The complete field of a general static spherically symmetric
  distribution of charge}.
\newblock {\em Nuovo Cimento}, A42:343--359, 1977.

\bibitem{Florides:1983}
P.~S.~{Florides}.
\newblock {The complete field of charged perfect fluid spheres and of other
  static spherically symmetric charged distributions}.
\newblock {\em Journal of Physics A: Mathematical General}, 16:1419--1433,
  1983.

\bibitem{Gron:1986}
{\O}.~{Gr{\o}n}.
\newblock {A charged generalization of Florides' interior Schwarzschild
  solution}.
\newblock {\em Gen. Rel. Grav.}, 18:591--596, 1986.

\bibitem{Kramer:1971}
D.~{Kramer} and G.~{Neugebauer}.
\newblock {Innere Reissner-Weyl-L{\"o}sung}.
\newblock {\em Annalen der Physik}, 482:129--135, 1971.

\bibitem{Mehra:1982}
A.~L.~{Mehra}.
\newblock {An interior solution for a charged sphere in general relativity}.
\newblock {\em Phys. Lett.}, A88:159--161, 1982.

\bibitem{Ortaggio:2002}
M.~Ortaggio.
\newblock Impulsive waves in the nariai universe.
\newblock {\em Phys. Rev.}, D65:084046, 2002.

\bibitem{Robinson:1959}
I.~Robinson.
\newblock A Solution of the Maxwell-Einstein Equations.
\newblock {\em Bull. Acad. Pol. Sci. Ser. Sci. Math. Astron. Phys.}, 7:351--352, 1959.

\bibitem{Wilson:1967}
S.~J.~Wilson.
\newblock Exact solution of a static charged sphere in general relativity.
\newblock {\em Can. J. Phys.}, 47:2401--2404, 1967.

\bibitem{Xu:1986ka}
C.-M.~Xu, X.-J.~Wu, and Z.~Huang.
\newblock A new class of spherically symmetric interior solution with
  cosmological constant lambda.
\newblock 1986.
\newblock IC-86-392.

\end{thebibliography}
\end{document}